\documentclass[11pt, a4paper]{article}
\usepackage{amsmath, fullpage}
\usepackage{sectsty}
\sectionfont{\fontsize{12}{12}\selectfont}
\subsectionfont{\fontsize{11}{12}\selectfont}
\usepackage{apacite}
\usepackage{amssymb}
\usepackage{graphicx}
\usepackage{authblk}
\newcommand\sbullet[1][.5]{\mathbin{\vcenter{\hbox{\scalebox{#1}{$\bullet$}}}}}

\usepackage{tikz}
\tikzset{every picture/.style={line width=0.75pt}}

\usepackage{lipsum}
\usepackage{titlesec}
\newtheorem{theorem}{Theorem}[section]

\usepackage{etoolbox}
\AtBeginEnvironment{theorem}{\small}

\usepackage{algorithm}
\usepackage{algpseudocode}

\titlespacing\section{0pt}{12pt plus 4pt minus 2pt}{0pt plus 2pt minus 2pt}
\titlespacing\subsection{0pt}{12pt plus 4pt minus 2pt}{0pt plus 2pt minus 2pt}

\begin{document}
\font\myfont=cmr12 at 15pt
\title{\myfont Continuous-Time Path-Dependent Exploratory Mean-Variance Portfolio Construction}
\author[1]{\small Zhou Fang}
\affil[1]{Department of Mathematics, The University of Texas at Austin}
\date{\small Feb 2023}
\maketitle

\begin{abstract}
    In this paper, we present an extended exploratory continuous-time mean-variance framework for portfolio management. Our strategy involves a new clustering method based on simulated annealing, which allows for more practical asset selection. Additionally, we consider past wealth evolution when constructing the mean-variance portfolio. We found that our strategy effectively learns from the past and performs well in practice.  
    
\end{abstract}

\section{Introduction}

The revolutionary work \cite{selection1952harry} is considered as the beginning of modern portfolio management, which proposes a framework for constructing portfolios in single periods. Much research in portfolio management came out after then. \cite{li2000optimal} considers the portfolio selection problem in multiple periods setting and \cite{zhou2000continuous} further studies the portfolio selection problem in a continuous-time setting, which makes portfolio management in high frequency possible. In \cite{wang2018exploration}, an innovative exploratory continuous-time mean-variance framework is first introduced. It replaces the deterministic policy with a stochastic policy to adjust assets' holdings, achieving the ideal trade-off between exploration and exploitation, and is considered more robust than the deterministic policy. This exploratory framework is further studied in a series of subsequent papers. \cite{wang2019large} extends the framework into a multi-assets setting. \cite{jia2022policy_a}, \cite{jia2022policy_b}, \cite{jia2022q} develops a more general framework for policy evaluation and policy improvement. 

In this paper, we further extend the exploratory continuous-time mean-variance framework by first clustering assets into several groups and then constructing a mean-variance portfolio by taking past wealth processes into consideration. In other words, there are phases in our framework. The first one is the clustering phase, and the second is the portfolio construction phase. The motivation for adding a clustering phase is to exclude similar assets and thus reduce the complexity. It is not possible and also not necessarily to hold thousands of stocks at one time for small retail and institutional investors. In the second portfolio construction phase, the motivation for considering the past wealth process or past performance is we think past experience should be important for making a current decision. When making a current decision, one should base it on past lessons together with the current wealth level instead of basing it only on the current wealth level. One intuitive example will be if a person came into the stock market in mid-2020, that person probably would think making money in the stock market is easy and maybe start to use high leverage in expect to make a higher gain, therefore, this person probably had a huge loss in 2022. However, if a person came into the stock market in 2008, that person probably would become more cautious, and therefore be aware of the market environment such as the Feds' interest rate policy. 

Many asset clustering methods have been developed in the past decades. As pointed out in \cite{tang2022asset}, one necessary criterion that wasn't in the previous literature is that correlations for two assets from the same cluster with any other asset should be similar. Following that paper's spirit, we use the similarity metric proposed in \cite{dolphin2021measuring} instead of using correlation. It is because correlation sometimes can't capture similarities between two assets, as pointed out in \cite{dolphin2021measuring} one asset can have negative returns, and the other asset can have high returns, but these two assets still are highly correlated. The similarity metric we use here takes cumulative returns into consideration and therefore can capture the pattern similarities between assets. The clustering method we use in this paper is a simulated-annealing-based method, which is inspired by \cite{ludkovski2022large}. We define an energy function that depends on asset similarities within a group. To our best knowledge, it is the first literature that uses this new similarity metric to cluster assets by a simulated annealing clustering method. From the empirical results, this method performs well, and can even possibly help investors find assets that guarantee statistical arbitrage opportunities, which requires further studies. 

There is a huge literature on portfolio management or portfolio optimization, but very few consider portfolio management problems in a path-dependent setting. The path-dependent setting in portfolio management means that the positions of assets should depend on the performance of past investments. Learning from the past is an important philosophy that shouldn't be ignored. The path-dependent case has technical difficulties, and there were no suitable tools for a long time. Thanks to the groundbreaking paper \cite{dupire2019functional} that introduces functional Ito calculus, which defines a new type of calculus on path space, and thus proposes a path version of the Ito formula, and the Feynmann-Kac formula. However, the path-dependent Hamilton-Jacobi-Bellman equation is very hard to solve numerically. Thanks to the recent development of deep learning, makes solving path-dependent Hamilton-Jacobi-Bellman equations possible. There are literature that uses neural networks to solve path-dependent PDE, such as \cite{saporito2020pdgm}, and \cite{sabate2020solving}. In this paper, we use PDGM to solve the PDE numerically, which is a neural network architecture whose main components are LSTM and feedforward networks. LSTM, as one of the most classical neural networks, has the ability to process sequential information and is naturally suitable to summarize past history. The feed-forward neural can then model certain functional by using the summaries given by LSTM. It is worth pointing out that LSTM is designed to forget some information in the past. In the deep learning community, the transformer basically replaced the LSTM to process sequential information. We believe replacing the LSTM component by a transformer component can have better results in solving path-dependent PDE, which requires further studies.

\section{Simulated Annealing Clustering}
Subsection 2.1 introduces a new similarity metric, and subsection 2.2 introduces the clustering method based on the proposed similarity metric via simulated annealing.
\subsection{\small Similarity Metric of Financial Time Series}
As pointed out in \cite{dolphin2021measuring}, two assets can be perfectly correlated, but their performances will be very different. It is because conventional correlation is a similarity m, correlation cannot be the only criterion to cluster assets with similar patterns assets, correlation cannot be the only criterion. In this paper, we use the metric proposed in \cite{dolphin2021measuring} to measure the similarity of two assets.

Assume there are two time-series $X = [x_1. x_2, ..., x_n]$, and $Y = [y_1, y_2, ..., y_n]$, which represent returns of two assets, and the similarity between these two time-series is defined to be  

\begin{equation}
\mathbf{sim}(X,Y) = \frac{w}{1 + e(X,Y)} + (1-w) \tau(X,Y), \hspace{0.2cm}  
\end{equation}

\begin{equation}
e(X,Y) = \sqrt{\big(\underset{1 \leq i \leq n}{\Pi}(1+x_i) - \underset{1 \leq i \leq n}{\Pi}(1+y_i)\big)^2}
\end{equation}
\begin{equation}
\tau(X,Y) = \frac{\underset{1 \leq i \leq n}{\sum} x_i y_i}{\sqrt{{\underset{1 \leq i \leq n}{\sum}} x_i^2}{\sqrt{\underset{1 \leq i \leq n}{\sum} y_i^2}}}
\end{equation}
In the above definition of similarity, $e(X,Y)$ measures the distance between the cumulative returns of two assets and $\tau(X,Y)$ is a modified version of correlation that measures the similarity of absolute performance instead of relative performance to mean returns. $w$ is a hyper-parameter to be determined. Empirically, the metric performs best when $w$ is between 0.4 and 0.6.

\subsection{\small Simulated Annealing Clustering}
The clustering procedure is based on simulated annealing. The key to the simulated annealing clustering method is the energy functions. The ideal energy function should control the size of each cluster, and let similar assets be in the same group. Let $C = \{C_1, C_2, ..., C_k\}$ be a partition of assets. Inspired by \cite{ludkovski2022large}, we propose a similar energy function
\begin{align}
E(C) = \underset{C_i \in C}{\sum} \Big( 1 - \frac{\kappa}{|C| - 1} \underset{X, Y \in C_i}{\sum} \mathbf{sim}(X,Y) \Big) 
\end{align}
In the above definition, $|C|$ is the number of clusters, which should never be 1, otherwise, it is not a clustering method. $\kappa$ is a hyper-parameter to be determined, which put a soft constraint on the number of clusters. 

The simulated annealing clustering method is as follows -- initially, all assets are in the same group, and for each iteration, a perturbation operation is applied to move one asset from one cluster to another with some criterion of accepting the perturbation. The criterion of accepting perturbation is whether or not the energy function is decreased or not. If the energy function is decreased, then one should accept the perturbation. If the energy function is increased, then one should accept the perturbation under a certain probability related to the energy function.

The exact simulated annealing clustering method is as follows \\
\textbf{Initialization:} Choose an initial temperature: $T_0$, final temperature: $T_f$, cooling rate: $\alpha < 1$, and let all assets in the same cluster. \\
\textbf{Step 1:} Assume current step is $l^{\text{th}}$ step. Denote the current partition as $C^{l}$, and apply perturbation operation to the current partition, which resulting a new partition, denote as $C^{l+1}_{new}$. Notice that the current temperature is $T_l$  \\
\textbf{Step 2:} $\Delta E^l = E(C^{l+1}_{new}) - E(C^l)$. \\
\indent \indent \textbf{If:} $\Delta E^l < 0$, accept partition $C^{l+1}_{new}$ as the partition $C^{l+1}$ for next iteration
\\
\indent \indent \textbf{Else:}
\begin{center}
    $C^{l+1}_{new}$ is accepted as $C^{l+1}$ with probability $\text{exp}(-\frac{\Delta E^l}{T_l})$ \\
    $C^{l}$ is accepted as $C^{l+1}$ with probability $1 - \text{exp}(-\frac{\Delta E^l}{T_l})$ 
\end{center} 
\textbf{Step 3:} Lowering the temperature based on cooling rate: $T_{l+1} = \alpha T_{l}$. 
\textbf{Step 4:} If $T_{l+1} < T_f$, end the clustering procedure. Otherwise, repeat steps 1 - 3                       

\section{Exploratory Mean-Variance Framework}
The exploratory mean-variance framework is first proposed in \cite{wang2018exploration} for one-dimensional asset dynamics and extends to multi-dimensional asset dynamics in \cite{wang2019large}. We follow \cite{wang2019large} to give an introduction to the exploratory mean-variance framework in this section.

The portfolio management problem is forming a portfolio that minimizes the variance under expected returns. 
\begin{align}
    \text{min }\mathbf{Var} (x_t) \\
    \text{s.t. }\mathbb{E}[x_t] = z
\end{align}
Assume the clustering phase is already completed, and there are $n$ clusters. We randomly select $n$ assets, by choosing one asset from each cluster, whose dynamics are denoted as $\{ S_t^1, S_t^2, ..., S_t^n \}$. For each asset $i$, assume that the underlying dynamics to be $$dS_t^i = S_t^i(\mu_t^i dt + \underset{1 \leq j \leq n}{\sum} \sigma_{t}^{ij} dW_t^j)$$
Let $\boldsymbol\mu_t = (\mu_t^1, \mu_t^2, ..., \mu_t^n)$, $\boldsymbol \sigma_t = (\sigma_t^{ij})$, and $ d \boldsymbol W_t = (dW_t^1, dW_t^2, ..., dW_t^n)$ denote the vector of drift rates, the covariance matrix, and the vector of n-dim independent Brownian motions respectively. Let $\mathbf{a_t} = (a_t^1, a_t^2, ..., a_t^n)$ denote the holdings (in dollar) of those $n$ assets at time $t$. Therefore, the wealth process will be 
\begin{equation}
    dx_t^{\mathbf{a_t}} = (\boldsymbol \mu_t - r\boldsymbol e_d)^{T} \boldsymbol a_t dt + \boldsymbol a_t^T \boldsymbol \sigma_t d\boldsymbol W_t
\end{equation}

In this paper, the holdings/control is path-dependent, which means holdings/control $\boldsymbol a_t$ depend on the past history of the wealth process. To differentiate the current wealth level and the path of the wealth process, we use lowercase letters for the current wealth level and use uppercase letters for the path of the wealth process. For example, $x_t^{\boldsymbol a_t}$ means the wealth level at time $t$ under holdings/control $\boldsymbol a_t$, while $X_t$ means a path of wealth process from time $0$ to time $t$. Now, the policy $ \boldsymbol \pi (\boldsymbol a_t | X_t)$ is a probability density, meaning that holdings/control at time $t$ has probability density $ \boldsymbol \pi (\boldsymbol a_t | X_t)$ to be at $\boldsymbol a_t$, based on a realized wealth process is $X_t$. 

Under the exploratory framework proposed in \cite{wang2018exploration}, we have a new stochastic process that denotes the average performance of the wealth process is 
\begin{equation}
{d\tilde{x}_t}^{\boldsymbol \pi} = (\boldsymbol \mu_t - r\boldsymbol e_d)^{T} \boldsymbol m_t dt + \boldsymbol m_t^T \boldsymbol \sigma_t d \boldsymbol W_t + \sqrt{\mathbf{Tr}(\boldsymbol \Sigma_t^{T} \boldsymbol C_t)}d \widetilde{W_t}
\end{equation}
Here, we assume that policy $\boldsymbol \pi$ is appied at time $t$. $\widetilde{W_t}$ is an independent Brownian motion, $\boldsymbol \Sigma_t = \boldsymbol \sigma_t \boldsymbol \sigma_t^{T}$ is a positive definite matrix. $\boldsymbol m_t$, and $\boldsymbol C_t$ are as follows, 
\begin{equation}
    \boldsymbol m_t = \int_{\mathbb{R}^n} \boldsymbol a_t \boldsymbol \pi(\boldsymbol a_t | \widetilde{X_t}) d \boldsymbol a_t
\end{equation}

\begin{equation}
    \boldsymbol C_t = \int_{\mathbb{R}^n} [\boldsymbol a_t - \boldsymbol m_t][\boldsymbol a_t - \boldsymbol m_t]^{T} \boldsymbol \pi(\boldsymbol a_t | \widetilde{ X_t}) d \boldsymbol a_t
\end{equation}
The goal is to identify a policy $\boldsymbol \pi$ that minimizes the following objective function, where $w$ is the Lagrangian multiplier
\begin{align} \mathbb{E}\Big[ (\tilde{x}_T^{\boldsymbol \pi} - w)^2 + \gamma \int_0^T \int_{\mathbb{R}^n} \boldsymbol \pi(\boldsymbol a_t | \widetilde{X_t}^{\boldsymbol \pi}) \log \boldsymbol \pi(\boldsymbol a_t | \widetilde{X_t}^{\boldsymbol \pi}) d \boldsymbol a_t dt  \Big] - (w-z)^2
\end{align}

This cost function is very similar to the cost function in \cite{wang2019large}, but instead, the cost function is path-dependent. The superscript $\boldsymbol \pi$ indicates wealth processes are generated under the policy $\boldsymbol \pi$. Now that we have path-dependent control and path-dependent cost function, we can't simply apply classical stochastic control results here, a functional version of stochastic control is needed.

\section{Path Dependent Stochastic Control}
\subsection{\small Functional Ito Calculus}
Functional Ito calculus is a crucial and powerful tool developed in \cite{dupire2019functional} to study path-dependent stochastic calculus problems. The following is a quick introduction to functional Ito calculus and some important consequences in the \cite{dupire2019functional}. 

Let $\Lambda_t$ denote the set of cadlag paths up to time $t$, and specifically, $\Lambda_t^n$ is the space of cadlag paths that takes values in $\mathbb{R}^n$. $\Lambda_t^{n \times k}$ denotes the space of two cadlag paths, one takes values in $\mathbb{R}^n$ and the other takes value in $\mathbb{R}^k$. In general, $\Lambda = \underset{t \in [0,T]}{\bigcup} \Lambda_t$, and $f: \Lambda \to \mathbb{R}^d$ is a functional.

If $X_t$ is a path that takes value in $\mathbb{R}$, and $X_t(u)$ indicates the value at time $u$. We define vertical, and flat extensions as follows, (For the path that takes value in $\mathbb{R}^n$, the above vertical and flat extensions are done on each dimension individually)
\[
  X_t^h(u) =
  \begin{cases}
        X_u & \text{if $0 \leq u < t$} \\
         X_t + h & \text{if $t = u$}
  \end{cases}
\]

\[
    X_{t, \delta t}(u) = 
    \begin{cases}
          X_u & \text{if $0 \leq u < t$} \\
          X_t & \text{if $t \leq u \leq t + \delta t$}
    \end{cases}
\]

Let $f: \Lambda \to \mathbb{R}$ be a functional, then the partial derivative as follows, (for $f: \Lambda \to \mathbb{R}^d$, the following definition is on each dimension individually)
$$\Delta_t f(X_t)= \underset{\delta t \to 0^{+}}{\text{lim}} \frac{f(X_{t, \delta t})  -  f(X_t)}{\delta t}$$
$$\Delta_x f(X_t) = \underset{h \to 0}{\text{inf}} \frac{f(X_t^h) - f(X_t)}{h}$$

To understand the functional Ito formula, we need to introduce the metric on the path space $\Lambda$. Let $X_t$, and $Y_s$ be two paths in $\Lambda$, and without loss of generality, assume $t \leq s$. Their distance is defined to be
\begin{equation}
    d_{\Lambda}(X_t, Y_s) = || X_{t,s-t} - Y_s||_{\infty} + (s-t)
\end{equation}
After defining the metric, we can define the continuity of a functional.

A functional $f: \Lambda \to \mathbb{R}$ is $\Lambda$-continuous at $X_t \in \Lambda$ if $\forall \epsilon > 0$, $\exists \delta > 0$, and $\forall Y_s$ such that $d_{\Lambda}(X_t, Y_s) < \delta$, there is $|f(X_t) - f(Y_s)| < \epsilon$. A functional is $\Lambda$-continuous if it is $\Lambda$-continuous at all paths. 

A functional $f: \Lambda \to \mathbb{R}$ is in $\mathbb{C}^{1, 2}$ if it is $\Lambda$-continuous, $C^2$ in x, and $C^1$ in t, with its partial derivatives also be $\Lambda$-continuous.

\begin{theorem}[Functional Ito Formula] 
Let $x_t$ be a continuous semi-martingale process, $f \in \mathbb{C}^{1,2}$, and $X_t$ is a path of process $x_t$. Then, for any $t \in [0,T]$,
\begin{align}
    f(X_t) = f(X_0) + \int_0^t \Delta_t f(X_s)ds + \int_0^t \Delta_x f(X_s)dx_s + \frac{1}{2} \int_0^t \Delta_{xx} f(X_s)d \langle x \rangle_s
\end{align}

\end{theorem}

\subsection{\small Functional Feynman-Kac Formula and Functional HJB equation}
Now, assume that $X_t$ is a path for the wealth process. If we use policy $\boldsymbol \pi$ for the left time, the cost functional are as follow,
\begin{equation}J(X_t, \boldsymbol \pi) = \mathbb{E} \Big[(\tilde{x}_T^{\boldsymbol \pi} - w)^2 + \gamma \int_t^T \int_{\mathbb{R}^n} \boldsymbol \pi(\boldsymbol a_s | \widetilde{X_s}^{\boldsymbol \pi}) \log \boldsymbol \pi(\boldsymbol a_s | \widetilde{X_s}^{\boldsymbol \pi}) d \boldsymbol a_s ds \hspace{0.1 cm} | \hspace{0.1cm} \widetilde{X_t}^{\boldsymbol \pi} = X_t \Big] - (w-z)^2 
\end{equation}
To simplify our computation, denote the following functional as
\begin{equation}
\widetilde{\boldsymbol b}(s, \widetilde{X_s}, \boldsymbol \pi) = (\boldsymbol \mu_s - r\boldsymbol e_d)^{T} \boldsymbol m_s 
\end{equation}
\begin{equation}
\widetilde{\boldsymbol \sigma}(s, \widetilde{X_s}, \boldsymbol \pi) = \big(\boldsymbol m_s^T \boldsymbol \sigma_s \hspace{0.1 cm}|\hspace{0.1cm} \sqrt{\mathbf{Tr}(\boldsymbol \Sigma_s^{T} \boldsymbol C_s)} \big)
\end{equation}
\begin{equation}
f(s, \widetilde{X_s}, \boldsymbol \pi) = \int_{\mathbb{R}^n} \boldsymbol \pi(\boldsymbol a_s | \widetilde{X_s}^{\boldsymbol \pi}) \log \boldsymbol \pi(\boldsymbol a_s | \widetilde{X_s}^{\boldsymbol \pi}) d \boldsymbol a_s
\end{equation}
Thus, the dynamics, and the cost functional are
\begin{equation}
d \widetilde{x_s}^{\boldsymbol \pi} = \widetilde{\boldsymbol b}(s, \widetilde{X_s}, \boldsymbol \pi) dt + \widetilde{\boldsymbol \sigma}(s, \widetilde{X_s}, \boldsymbol \pi) (d \boldsymbol W_t \hspace{0.1cm} | \hspace{0.1cm} d \widetilde{W_t})^{T}
\end{equation}
\begin{equation}
J(X_t, \boldsymbol \pi) = \mathbb{E} \Big[ (\tilde{x}_T^{\boldsymbol \pi} - w)^2 + \gamma \int_t^T f(s, \widetilde{X_s}, \boldsymbol \pi) ds  \hspace{0.1 cm} | \hspace{0.1cm} \widetilde{X_t}^{\boldsymbol \pi} = X_t \Big] - (w-z)^2
\end{equation}
Let $\tau > t$, and consider the following,  
\begin{equation}
Y_{\tau} = J(\widetilde{X_{\tau}}, \boldsymbol \pi) + \gamma \int_t^{\tau} f(s,\widetilde{X_s}, \boldsymbol \pi)ds
\end{equation}
Consider the following, 
\begin{align*}
    \mathbb{E}[Y_\tau | X_t] &=  \mathbb{E}\Big[ \mathbb{E}[  (\tilde{x}_T^{\pi} - w)^2 + \gamma \int_\tau^T f(s,\widetilde{X_s}, \boldsymbol \pi)ds  | \widetilde{X_\tau} ]   + \int_t^{\tau}  f(s,\widetilde{X_s}, \boldsymbol \pi)ds \Big| X_t \Big] - (w-z)^2 \\
    &= \mathbb{E}\Big[ \mathbb{E}\big[  (\tilde{x}_T^{\pi} - w)^2 + \gamma \int_\tau^T f(s,\widetilde{X_s}, \boldsymbol \pi)ds  \big| \widetilde{X_\tau} \big] \Big| X_t \Big] + \mathbb{E}\Big[\int_t^{\tau} f(s,\widetilde{X_s}, \boldsymbol \pi)ds \Big| X_t \Big] - (w-z)^2 \\
    &= \mathbb{E}\Big[  (\tilde{x}_T^{\pi} - w)^2 + \gamma \int_\tau^T f(s,\widetilde{X_s}, \boldsymbol \pi)ds  \Big | X_t  \Big] + \mathbb{E}\Big[\int_t^{\tau} f(s,\widetilde{X_s}, \boldsymbol \pi)ds \Big| X_t \Big] - (w-z)^2 \\ 
    &= \mathbb{E}\Big[  (\tilde{x}_T^{\pi} - w)^2 + \gamma \int_t^T f(s,\widetilde{X_s}, \boldsymbol \pi)ds  \Big | X_t  \Big] - (w-z)^2 \\
    &= J(X_t, \pi) \\
    &= Y_t
\end{align*}
Therefore, we can see the functional $Y_{\tau}$ is martingale, by the functional Ito formula, the following can be derived
\begin{equation}
\widetilde{\boldsymbol b}(\tau, \widetilde{X_{\tau}}, \boldsymbol \pi)\Delta_x J(\widetilde{X_{\tau}}, \boldsymbol \pi)  + \Delta_{t} J(\widetilde{X_{\tau}}, \boldsymbol \pi) + \frac{1}{2} \widetilde{\boldsymbol \sigma}(\tau, \widetilde{X_{\tau}}, \boldsymbol \pi) \widetilde{\boldsymbol \sigma}(\tau, \widetilde{X_{\tau}}, \boldsymbol \pi)^{T} \Delta_{xx} J(\widetilde{X_{\tau}}, \boldsymbol \pi) + \gamma f(\tau, \widetilde{X_{\tau}}, \boldsymbol \pi) = 0
\end{equation}
To simplify the notation, assume the policy $\boldsymbol \pi$ be applied at time $0$, then at time $t$, and there is a path $X_t = \widetilde{X}^{\boldsymbol \pi}_t$.  
, and we have functional Feynman-Kac formula,
\begin{equation} \widetilde{\boldsymbol b}(t, X_t, \boldsymbol \pi) \Delta_x J(X_{t}, \boldsymbol \pi)  + \Delta_{t} J(X_{t}, \boldsymbol \pi) + \frac{1}{2} \widetilde{\boldsymbol \sigma}(t, X_t, \boldsymbol \pi) \widetilde{\boldsymbol \sigma}(t, X_t, \boldsymbol \pi)^{T} \Delta_{xx} J(X_{t}, \boldsymbol \pi) + \gamma f(t, X_{t}, \boldsymbol \pi) = 0
\end{equation}
Now, let $V(X_t) = \underset{\boldsymbol \pi}{\text{inf}} \hspace{0.1cm} J(X_t, \boldsymbol \pi)$, the following is an informal derivation of path-dependent HJB equation. The rigorous derivation of path-dependent HJB equation for portfolio management problems faces lots of technical challenges, which is beyond the scope of this paper.
\begin{align}
    V(X_t) &= \underset{\pi}{\text{inf}} \hspace{0.1cm} J(X_t, \boldsymbol \pi) \\
    &= \underset{\pi}{\text{inf}} \hspace{0.1cm} \mathbb{E}\Big[  (\tilde{x}_T^{\pi} - w)^2 + \gamma \int_t^T f(s,\widetilde{X_s}, \boldsymbol \pi)ds  \Big | X_t  \Big] - (w-z)^2 \\
    &= \underset{\pi}{\text{inf}} \hspace{0.1cm} \mathbb{E}\Big[ \mathbb{E}\big[  (\tilde{x}_T^{\pi} - w)^2 + \gamma \int_{t+dt}^T f(s,\widetilde{X_s}, \boldsymbol \pi)ds  \big| \widetilde{X}_{t+dt} \big] \Big| X_t \Big] - (w-z)^2 + \mathbb{E}\Big[   \gamma \int_t^{t+dt}   f(s,\widetilde{X_s}, \boldsymbol \pi)ds  \Big| X_t    \Big] \\
    &= \underset{\pi}{\text{inf}} \hspace{0.1cm} \mathbb{E}\Big[  (\tilde{x}_T^{\pi} - w)^2 + \gamma \int_t^T f(s,\widetilde{X_s}, \boldsymbol \pi)ds  \Big | X_t  \Big] - (w-z)^2 \\
    &= \underset{\pi}{\text{inf}} \hspace{0.1cm} \mathbb{E}\Big[ \mathbb{E}\big[  (\tilde{x}_T^{\pi} - w)^2 + \gamma \int_{t+dt}^T f(s,\widetilde{X_s}, \boldsymbol \pi)ds  \big| \widetilde{X}_{t+dt} \big] - (w-z)^2 \Big| X_t \Big]  + \mathbb{E}\Big[   \gamma \int_t^{t+dt}   f(s,\widetilde{X_s}, \boldsymbol \pi)ds  \Big| X_t    \Big] 
\end{align}
Since the first term in the above equation depends on the policy applied from time $t+dt$, which is independent from the policy applied during the interval $[t, t+dt]$. In other words, the policy $\pi$ appears in the above equation is a combination of policy applied in the interval $[t, t+dt]$, and the policy applied after time $t+dt$. Therefore, the above equation becomes
\begin{equation}
    V(X_t)  = \underset{\pi}{\text{inf}} \hspace{0.1cm} \mathbb{E}\big[  V(X_{t+dt}) \big| X_t \big] + \mathbb{E}\big[   \gamma \int_t^{t+dt}   f(s,\widetilde{X_s}, \boldsymbol \pi)ds  \big| X_t  \big] 
\end{equation}
Moving the left-hand side to the right, and letting $dt \to 0$, the above equation becomes, 
\begin{align}
    0 &= \underset{\pi}{\text{inf}} \hspace{0.1cm} \mathbb{E}\big[  V(X_{t+dt}) - V(X_t) \big| X_t \big] + \mathbb{E}\big[   \gamma \int_t^{t+dt}   f(s,\widetilde{X_s}, \boldsymbol \pi)ds  \big| X_t  \big] \\ &= \underset{\pi}{\text{inf}} \hspace{0.1cm} \mathbb{E}\big[ \Delta_t V(X_t) + \Delta_x V(X_t) d\tilde{x}_t^{\boldsymbol \pi} + \frac{1}{2} \Delta_{xx} V(X_t) d \langle \tilde{x}^{\boldsymbol \pi}\rangle_t  \big| X_t \big] + \gamma f(t, X_t, \boldsymbol \pi) dt \\
    &= \underset{\pi}{\text{inf}} \hspace{0.1cm} \big [\Delta_t V(X_t) + \widetilde{\boldsymbol b}(t, X_t, \boldsymbol \pi) \Delta_x V(X_t) + \frac{1}{2} \widetilde{\boldsymbol \sigma}(t, X_t, \boldsymbol \pi) \widetilde{\boldsymbol \sigma}(t, X_t, \boldsymbol \pi)^{T} \Delta_{xx} V(X_t) + \gamma f(t, X_t, \boldsymbol \pi)  \big ] dt
\end{align}
Divide both sides by $dt$, we have the following path-dependent HJB equation
\begin{equation}
\Delta_t V(X_t) + \underset{\boldsymbol \pi}{\text{inf}} \hspace{0.1cm} \big \{ \widetilde{\boldsymbol b}(t, X_t, \boldsymbol \pi) \Delta_x V(X_t) + \frac{1}{2} \widetilde{\boldsymbol \sigma}(t, X_t, \boldsymbol \pi) \widetilde{\boldsymbol \sigma}(t, X_t, \boldsymbol \pi)^{T} \Delta_{xx} V(X_t) + \gamma f(t, X_t, \boldsymbol \pi)  \big \}  = 0
\end{equation}
Consider the formula inside the $inf$, plug in $\widetilde{\boldsymbol b}(t, X_t, \boldsymbol \pi)$, $\widetilde{\boldsymbol \sigma}(t, X_t, \boldsymbol \pi)$, and $f(t, X_t, \boldsymbol \pi)$, we have the following formula, 

\begin{equation}
\int_{\mathbb{R}^n} \Big[ (\boldsymbol \mu_t - r \boldsymbol e_d)^{T} \boldsymbol a_t \Delta_x V(X_t) + \frac{1}{2}\boldsymbol a_t^{T} \boldsymbol \sigma_t \boldsymbol \sigma_t^{T} \boldsymbol a_t \Delta_{xx} V(X_t) + \gamma \log \boldsymbol \pi(\boldsymbol a_t | X_t) \Big ] \boldsymbol \pi(\boldsymbol a_t | X_t) d \boldsymbol a_t
\end{equation}
To find a candidate for the optimal policy, the following equation has to be independent of $\boldsymbol a_t$ and only depend on $X_t$
\begin{equation}
L(\boldsymbol a_t, X_t, \boldsymbol \pi) = (\boldsymbol \mu_t - r \boldsymbol e_d)^{T} \boldsymbol a_t \Delta_x V(X_t) + \frac{1}{2} \boldsymbol a_t^{T} \boldsymbol \sigma_t \boldsymbol \sigma_t^{T} \boldsymbol a_t \Delta_{xx} V(X_t) + \gamma \log \boldsymbol \pi(\boldsymbol a_t | X_t)
\end{equation}
\begin{equation}
    U(\boldsymbol a_t, X_t) = (\boldsymbol \mu_t - r \boldsymbol e_d)^{T} \boldsymbol a_t \Delta_x V(X_t) + \frac{1}{2} \boldsymbol a_t^{T} \boldsymbol \sigma_t \boldsymbol \sigma_t^{T} \boldsymbol a_t \Delta_{xx} V(X_t)
\end{equation}
Assume $L(\boldsymbol a_t, X_t, \boldsymbol \pi)$ depend on the $\boldsymbol a_t$, and by the fact that $L(\boldsymbol a_t, X_t, \boldsymbol \pi)$ is continuous at every point $\boldsymbol a_t$, then without loss of generality, there exist two regions $V_1$, and $V_2$ such that $L(\boldsymbol a_t, X_t, \boldsymbol \pi) - L(\boldsymbol a'_t, X_t, \boldsymbol \pi) > h$ for points $\boldsymbol a_t \in V_1$, and $\boldsymbol a'_t \in V_2$. Without loss of generality, assume the volume of these two regions is equal to $V$. Now, let $\epsilon$ be sufficiently small, and consider the upgraded $\boldsymbol \pi^{*}$ that $\boldsymbol \pi^{*}(\boldsymbol a_t | X_t) = \boldsymbol \pi(\boldsymbol a_t | X_t) + \epsilon$ for $\boldsymbol a_t \in V_1$, and $\boldsymbol \pi^{*}(\boldsymbol a_t | X_t) = \boldsymbol \pi(\boldsymbol a_t | X_t) - \epsilon$ for $\boldsymbol a_t \in V_2$, and $\boldsymbol \pi^{*}$ is same for the rest space. Therefore, the change of $L(\boldsymbol a_t, X_t, \boldsymbol \pi) \boldsymbol \pi(\boldsymbol a_t | X_t)$ for regions $V_1$ is
\begin{align}
    & L(\boldsymbol a_t, X_t, \boldsymbol \pi^{*}) \boldsymbol \pi^{*}(\boldsymbol a_t | X_t) - L(\boldsymbol a_t, X_t, \boldsymbol \pi) \boldsymbol \pi(\boldsymbol a_t | X_t) \\
    =& \epsilon U(\boldsymbol a_t, X_t) + \gamma (\boldsymbol \pi(\boldsymbol a_t | X_t) + \epsilon) \log(\boldsymbol \pi(\boldsymbol a_t | X_t) + \epsilon) - \gamma \boldsymbol \pi (\boldsymbol a_t | X_t) \log\boldsymbol \pi (\boldsymbol a_t | X_t) \\
    =& \epsilon U(\boldsymbol a_t, X_t) +  \gamma (\boldsymbol \pi(\boldsymbol a_t | X_t) + \epsilon) \big [ \log \boldsymbol \pi(\boldsymbol a_t | X_t) +  \log (1 + \frac{\epsilon}{\boldsymbol \pi(\boldsymbol a_t | X_t)})  \big ] - \gamma \boldsymbol \pi (\boldsymbol a_t | X_t) \log\boldsymbol \pi (\boldsymbol a_t | X_t) \\
    =& \epsilon U(\boldsymbol a_t, X_t) + \epsilon \gamma \log\boldsymbol \pi (\boldsymbol a_t | X_t) + \gamma \epsilon + \mathcal{O}(\epsilon^2)
\end{align}
Here, since the $\epsilon$ is sufficiently small, we use Taylor expansion for the $\log (1 + \frac{\epsilon}{\boldsymbol \pi(\boldsymbol a_t | X_t)})$ to the (37). Similarly, for region $V_2$, the change of the $L(\boldsymbol a_t, X_t, \boldsymbol \pi) \boldsymbol \pi(\boldsymbol a_t | X_t)$ is 
\begin{align}
    & L(\boldsymbol a_t, X_t, \boldsymbol \pi^{*}) \boldsymbol \pi^{*}(\boldsymbol a_t | X_t) - L(\boldsymbol a_t, X_t, \boldsymbol \pi) \boldsymbol \pi(\boldsymbol a_t | X_t) \\
    =& -\epsilon U(\boldsymbol a_t, X_t) + \gamma (\boldsymbol \pi(\boldsymbol a_t | X_t) - \epsilon) \log(\boldsymbol \pi(\boldsymbol a_t | X_t) - \epsilon) - \gamma \boldsymbol \pi (\boldsymbol a_t | X_t) \log\boldsymbol \pi (\boldsymbol a_t | X_t) \\
    =& -\epsilon U(\boldsymbol a_t, X_t) +  \gamma (\boldsymbol \pi(\boldsymbol a_t | X_t) - \epsilon) \big [ \log \boldsymbol \pi(\boldsymbol a_t | X_t) +  \log (1 - \frac{\epsilon}{\boldsymbol \pi(\boldsymbol a_t | X_t)})  \big ] - \gamma \boldsymbol \pi (\boldsymbol a_t | X_t) \log\boldsymbol \pi (\boldsymbol a_t | X_t) \\
    =& -\epsilon U(\boldsymbol a_t, X_t) - \epsilon \gamma \log\boldsymbol \pi (\boldsymbol a_t | X_t) - \gamma \epsilon + \mathcal{O}(\epsilon^2)
\end{align}
Therefore, the change of (31) is
\begin{align}
    &\int_{V_1} \big[ \epsilon U(\boldsymbol a_t, X_t) + \epsilon \gamma \log\boldsymbol \pi (\boldsymbol a_t | X_t) + \gamma \epsilon \big] d \boldsymbol a_t - \int_{V_2} \big[\epsilon U(\boldsymbol a_t, X_t) + \epsilon \gamma \log\boldsymbol \pi (\boldsymbol a_t | X_t) + \gamma \epsilon \big] d\boldsymbol a_t  \\
    =& \epsilon \int_{V_1}  U(\boldsymbol a_t, X_t) + \gamma \log\boldsymbol \pi (\boldsymbol a_t | X_t) d \boldsymbol a_t - \epsilon \int_{V_2}  U(\boldsymbol a_t, X_t) + \gamma \log\boldsymbol \pi (\boldsymbol a_t | X_t) d \boldsymbol a_t + \epsilon \gamma \int_{V_1} d \boldsymbol a_t - \epsilon \gamma \int_{V_2} d \boldsymbol a_t \\
    =&\epsilon \int_{V_1}  U(\boldsymbol a_t, X_t) + \gamma \log\boldsymbol \pi (\boldsymbol a_t | X_t) d \boldsymbol a_t - \epsilon \int_{V_2}  U(\boldsymbol a_t, X_t) + \gamma \log\boldsymbol \pi (\boldsymbol a_t | X_t) d \boldsymbol a_t + V - V \\
    =& \epsilon \int_{V_1} L(\boldsymbol a_t, X_t, \boldsymbol \pi) d \boldsymbol a_t - \epsilon \int_{V_2} L(\boldsymbol a_t, X_t, \boldsymbol \pi) d \boldsymbol a_t > \epsilon h V 
\end{align}
From the above derivation, one can see that if the policy depends on $\boldsymbol a_t$, there is always an improved policy by slightly adjusting the probability density.  
\begin{equation}(\boldsymbol \mu_t - r \boldsymbol e_d)^{T} \boldsymbol a_t \Delta_x V(X_t) + \frac{1}{2} \boldsymbol a_t^{T} \boldsymbol \sigma_t \boldsymbol \sigma_t^{T} \boldsymbol a_t \Delta_{xx} V(X_t) + \gamma \log \boldsymbol \pi(\boldsymbol a_t | X_t) = C(X_t)
\end{equation}
Thus, the optimal policy is of the following form, where $A$ is constant
\begin{equation}
\boldsymbol \pi(\boldsymbol a_t | X_t)  = A(X_t) \hspace{0.1cm} \text{exp}\Big(- \frac{1}{\gamma}\big( (\boldsymbol \mu_t - r \boldsymbol e_d)^{T} \boldsymbol a_t \Delta_x V(X_t) + \frac{1}{2} \boldsymbol a_t^{T} \boldsymbol \sigma_t \boldsymbol \sigma_t^{T} \boldsymbol a_t \Delta_{xx} V(X_t)\big) \Big)
\end{equation}
By the fact that $\boldsymbol \pi(\boldsymbol a_t | X_t)$ is a probability distribution, we have
\begin{equation} \boldsymbol \pi(\boldsymbol a_t | X_t) = \frac{ \text{exp}\Big(- \frac{1}{\gamma}\big( (\boldsymbol \mu_t - r \boldsymbol e_d)^{T} \boldsymbol a_t \Delta_x V(X_t) + \frac{1}{2} \boldsymbol a_t^{T} \boldsymbol \sigma_t \boldsymbol \sigma_t^{T} \boldsymbol a_t \Delta_{xx} V(X_t)\big) \Big)}{\int_{\mathbb{R}^n} \text{exp}\Big(- \frac{1}{\gamma}\big( (\boldsymbol \mu_t - r \boldsymbol e_d)^{T} \boldsymbol a_t \Delta_x V(X_t) + \frac{1}{2} \boldsymbol a_t^{T} \boldsymbol \sigma_t \boldsymbol \sigma_t^{T} \boldsymbol a_t \Delta_{xx} V(X_t)\big) \Big) d \boldsymbol a_t}
\end{equation}
Thus, $\boldsymbol \pi (\boldsymbol a_t | X_t)$ is Gaussian distribution, more specifically, 
\begin{equation}
    \boldsymbol \pi ( \hspace{0.1cm} \sbullet \hspace{0.1cm} | X_t) \sim \mathcal{N}\Big( - (\boldsymbol \sigma_t \boldsymbol \sigma_t^{T})^{-1} (\boldsymbol \mu_t - r \boldsymbol e_d)\frac{\Delta_x V(X_t)}{\Delta_{xx} V(X_t)} \hspace{0.1cm}, \hspace{0.1cm}  (\boldsymbol \sigma_t \boldsymbol \sigma_t^{T})^{-1} \frac{\gamma}{\Delta_{xx} V(X_t)} \Big ) 
\end{equation}
Plug the policy back into HJB equation, as shown in \cite{wang2019large}, and the HJB equation becomes,
\begin{equation}
    \Delta_t V(X_t) - \frac{|| \boldsymbol \sigma_t^{-1} (\boldsymbol \mu_t - r \boldsymbol e_d) ||^2 }{2} \frac{[\Delta_x V(X_t)]^2}{\Delta_{xx} V(X_t)} + \frac{\gamma}{2} \bigg [  d - d \hspace{0.1cm} \text{ln } \frac{2 \pi e \gamma}{\Delta_{xx} V(X_t)} + \text{ln } \text{det} (\boldsymbol \sigma_t^{T} \boldsymbol \sigma_t)  \bigg ] = 0
\end{equation}

In path-dependent cases, it is impossible to write an analytic solution for equation $(28)$. Solving equation $(28)$ numerically is the only way and should be sufficient for practitioners. It is natural to solve path-dependent PDE numerically using neural networks because the deep learning community has developed ways to deal with sequential information. 

\section{Deep Learning PDE Solver}
To solve $(28)$ numerically, it is important to approximate the value function, $V(X_t)$, where $X_t$ is a path of wealth process up to time $t$. However, there is no magic in deep learning. None of the deep learning algorithms can directly deal with continuous sample paths and output values. Therefore, we have to discretize the time and sample path, plug the discretized version into a neural network, and get the output value. 

Perhaps the most famous neural network to capture long-term dependence is Long Short-Term Memory (LSTM), which is a special type of recurrent neural network but doesn't successfully mitigate the problem of gradient exploding and gradient vanishing (Those two are the same thing). Plugging our discretized version of the sample path into LSTM, we can get outputs that represent the path information in a neat way. After capturing the path information, there is one more thing to do -- modeling the value function based on those outputs that contain path information. It is an easier step, because of the universal approximation theorem, we can just use a feed-forward neural network. In the rest of this section, we follow the framework proposed in \cite{saporito2020pdgm} to solve our equation. 

\subsection{Long Short-Term Memory}
First, LSTM is not a specific neural network structure, but more like a building block instead. The structure of an LSTM block is as follows, 

\begin{center}
    
\begin{tikzpicture}[x=0.75pt,y=0.75pt,yscale=-1,xscale=1]
%uncomment if require: \path (0,439); %set diagram left start at 0, and has height of 439

%Shape: Rectangle [id:dp4915959611200651] 
\draw   (200,150) -- (279.52,150) -- (279.52,190) -- (200,190) -- cycle ;
%Straight Lines [id:da2630655365622079] 
\draw    (160.52,160) -- (197.52,160) ;
\draw [shift={(199.52,160)}, rotate = 180] [color={rgb, 255:red, 0; green, 0; blue, 0 }  ][line width=0.75]    (10.93,-3.29) .. controls (6.95,-1.4) and (3.31,-0.3) .. (0,0) .. controls (3.31,0.3) and (6.95,1.4) .. (10.93,3.29)   ;
%Straight Lines [id:da2198228893803582] 
\draw    (160.52,180) -- (197.52,180) ;
\draw [shift={(199.52,180)}, rotate = 180] [color={rgb, 255:red, 0; green, 0; blue, 0 }  ][line width=0.75]    (10.93,-3.29) .. controls (6.95,-1.4) and (3.31,-0.3) .. (0,0) .. controls (3.31,0.3) and (6.95,1.4) .. (10.93,3.29)   ;
%Straight Lines [id:da6916230958539553] 
\draw    (279.52,180) -- (316.52,180) ;
\draw [shift={(318.52,180)}, rotate = 180] [color={rgb, 255:red, 0; green, 0; blue, 0 }  ][line width=0.75]    (10.93,-3.29) .. controls (6.95,-1.4) and (3.31,-0.3) .. (0,0) .. controls (3.31,0.3) and (6.95,1.4) .. (10.93,3.29)   ;
%Straight Lines [id:da1032352528991014] 
\draw    (279.52,160) -- (316.52,160) ;
\draw [shift={(318.52,160)}, rotate = 180] [color={rgb, 255:red, 0; green, 0; blue, 0 }  ][line width=0.75]    (10.93,-3.29) .. controls (6.95,-1.4) and (3.31,-0.3) .. (0,0) .. controls (3.31,0.3) and (6.95,1.4) .. (10.93,3.29)   ;
%Straight Lines [id:da22442082381635142] 
\draw    (240.52,210) -- (240.52,192) ;
\draw [shift={(240.52,190)}, rotate = 90] [color={rgb, 255:red, 0; green, 0; blue, 0 }  ][line width=0.75]    (10.93,-3.29) .. controls (6.95,-1.4) and (3.31,-0.3) .. (0,0) .. controls (3.31,0.3) and (6.95,1.4) .. (10.93,3.29)   ;

% Text Node
\draw (131,156) node [anchor=north west][inner sep=0.75pt]  [font=\scriptsize] [align=left] {a\textsubscript{$\displaystyle i-1$}};
% Text Node
\draw (132,175) node [anchor=north west][inner sep=0.75pt]  [font=\scriptsize] [align=left] {c\textsubscript{$\displaystyle i-1$}};
% Text Node
\draw (323,157) node [anchor=north west][inner sep=0.75pt]  [font=\scriptsize] [align=left] {a\textsubscript{i}};
% Text Node
\draw (322,177) node [anchor=north west][inner sep=0.75pt]  [font=\scriptsize] [align=left] {c\textsubscript{i}};
% Text Node
\draw (221,162) node [anchor=north west][inner sep=0.75pt]   [align=left] {LSTM};
% Text Node
\draw (236,212) node [anchor=north west][inner sep=0.75pt]  [font=\scriptsize] [align=left] {x\textsubscript{i}};

\end{tikzpicture}

\end{center}

In the above block diagram, there are three inputs, $a_{i - 1}$, $c_{i - 1}$, and $x_{i}$, and there are two outputs $a_i$, and $c_i$. Intuitively, one can think in this way, $a_{i-1}$, as a single number, is a neat way to represent all the information for the first $i - 1$ steps, so people call it "long-term memory". While for $c_{i-1}$, the short-term memory, as the name indicated, recent several steps' information has more influence on it. At last, $x_i$ represents the information one gets from the current step. Those three memories interact with each other to generate new long-term memory and short-term memory, $a_i$, and $c_i$. 

Back to our equation, assume now we have a sample path of the wealth process $X_t$, where $t \in [0, T]$. We discretize the time interval into $N$ equal periods, $0 = t_0 < t_1 < ... < t_N = T$, and $\delta t = t_{i+1} - t_i$, which will be useful later. Let $x_i$, where $i = 0, 1, 2, ..., N$ be the wealth at time $t_i$. So, we have the following network that consist of $N+1$ blocks of LSTM. Here, $a_{-1} = 0$, and $c_{-1} = 0$ represent there is no memory initially.

\begin{center}
\begin{tikzpicture}[x=0.75pt,y=0.75pt,yscale=-1,xscale=1]
%uncomment if require: \path (0,439); %set diagram left start at 0, and has height of 439

%Shape: Rectangle [id:dp4915959611200651] 
\draw   (200,150) -- (279.52,150) -- (279.52,190) -- (200,190) -- cycle ;
%Straight Lines [id:da2630655365622079] 
\draw    (160.52,160) -- (197.52,160) ;
\draw [shift={(199.52,160)}, rotate = 180] [color={rgb, 255:red, 0; green, 0; blue, 0 }  ][line width=0.75]    (10.93,-3.29) .. controls (6.95,-1.4) and (3.31,-0.3) .. (0,0) .. controls (3.31,0.3) and (6.95,1.4) .. (10.93,3.29)   ;
%Straight Lines [id:da2198228893803582] 
\draw    (160.52,180) -- (197.52,180) ;
\draw [shift={(199.52,180)}, rotate = 180] [color={rgb, 255:red, 0; green, 0; blue, 0 }  ][line width=0.75]    (10.93,-3.29) .. controls (6.95,-1.4) and (3.31,-0.3) .. (0,0) .. controls (3.31,0.3) and (6.95,1.4) .. (10.93,3.29)   ;
%Straight Lines [id:da6916230958539553] 
\draw    (279.52,180) -- (316.52,180) ;
\draw [shift={(318.52,180)}, rotate = 180] [color={rgb, 255:red, 0; green, 0; blue, 0 }  ][line width=0.75]    (10.93,-3.29) .. controls (6.95,-1.4) and (3.31,-0.3) .. (0,0) .. controls (3.31,0.3) and (6.95,1.4) .. (10.93,3.29)   ;
%Straight Lines [id:da1032352528991014] 
\draw    (279.52,160) -- (316.52,160) ;
\draw [shift={(318.52,160)}, rotate = 180] [color={rgb, 255:red, 0; green, 0; blue, 0 }  ][line width=0.75]    (10.93,-3.29) .. controls (6.95,-1.4) and (3.31,-0.3) .. (0,0) .. controls (3.31,0.3) and (6.95,1.4) .. (10.93,3.29)   ;
%Straight Lines [id:da22442082381635142] 
\draw    (240.52,210) -- (240.52,192) ;
\draw [shift={(240.52,190)}, rotate = 90] [color={rgb, 255:red, 0; green, 0; blue, 0 }  ][line width=0.75]    (10.93,-3.29) .. controls (6.95,-1.4) and (3.31,-0.3) .. (0,0) .. controls (3.31,0.3) and (6.95,1.4) .. (10.93,3.29)   ;
%Shape: Rectangle [id:dp9861771050557273] 
\draw   (580,150) -- (659.52,150) -- (659.52,190) -- (580,190) -- cycle ;
%Straight Lines [id:da7642681575374748] 
\draw    (541.52,159) -- (578.52,159) ;
\draw [shift={(580.52,159)}, rotate = 180] [color={rgb, 255:red, 0; green, 0; blue, 0 }  ][line width=0.75]    (10.93,-3.29) .. controls (6.95,-1.4) and (3.31,-0.3) .. (0,0) .. controls (3.31,0.3) and (6.95,1.4) .. (10.93,3.29)   ;
%Straight Lines [id:da6466370347681569] 
\draw    (540.52,180) -- (577.52,180) ;
\draw [shift={(579.52,180)}, rotate = 180] [color={rgb, 255:red, 0; green, 0; blue, 0 }  ][line width=0.75]    (10.93,-3.29) .. controls (6.95,-1.4) and (3.31,-0.3) .. (0,0) .. controls (3.31,0.3) and (6.95,1.4) .. (10.93,3.29)   ;
%Straight Lines [id:da8563077138312174] 
\draw    (659.52,160) -- (696.52,160) ;
\draw [shift={(698.52,160)}, rotate = 180] [color={rgb, 255:red, 0; green, 0; blue, 0 }  ][line width=0.75]    (10.93,-3.29) .. controls (6.95,-1.4) and (3.31,-0.3) .. (0,0) .. controls (3.31,0.3) and (6.95,1.4) .. (10.93,3.29)   ;
%Straight Lines [id:da46164119642863843] 
\draw    (659.52,180) -- (696.52,180) ;
\draw [shift={(698.52,180)}, rotate = 180] [color={rgb, 255:red, 0; green, 0; blue, 0 }  ][line width=0.75]    (10.93,-3.29) .. controls (6.95,-1.4) and (3.31,-0.3) .. (0,0) .. controls (3.31,0.3) and (6.95,1.4) .. (10.93,3.29)   ;
%Shape: Rectangle [id:dp9407751896600098] 
\draw   (381,150) -- (460.52,150) -- (460.52,190) -- (381,190) -- cycle ;
%Straight Lines [id:da7534856042930864] 
\draw    (340.52,160) -- (377.52,160) ;
\draw [shift={(379.52,160)}, rotate = 180] [color={rgb, 255:red, 0; green, 0; blue, 0 }  ][line width=0.75]    (10.93,-3.29) .. controls (6.95,-1.4) and (3.31,-0.3) .. (0,0) .. controls (3.31,0.3) and (6.95,1.4) .. (10.93,3.29)   ;
%Straight Lines [id:da9868958474000051] 
\draw    (340.52,180) -- (377.52,180) ;
\draw [shift={(379.52,180)}, rotate = 180] [color={rgb, 255:red, 0; green, 0; blue, 0 }  ][line width=0.75]    (10.93,-3.29) .. controls (6.95,-1.4) and (3.31,-0.3) .. (0,0) .. controls (3.31,0.3) and (6.95,1.4) .. (10.93,3.29)   ;
%Straight Lines [id:da7456991088827707] 
\draw    (461.52,181) -- (498.52,181) ;
\draw [shift={(500.52,181)}, rotate = 180] [color={rgb, 255:red, 0; green, 0; blue, 0 }  ][line width=0.75]    (10.93,-3.29) .. controls (6.95,-1.4) and (3.31,-0.3) .. (0,0) .. controls (3.31,0.3) and (6.95,1.4) .. (10.93,3.29)   ;
%Straight Lines [id:da5341906298481356] 
\draw    (461.52,159) -- (498.52,159) ;
\draw [shift={(500.52,159)}, rotate = 180] [color={rgb, 255:red, 0; green, 0; blue, 0 }  ][line width=0.75]    (10.93,-3.29) .. controls (6.95,-1.4) and (3.31,-0.3) .. (0,0) .. controls (3.31,0.3) and (6.95,1.4) .. (10.93,3.29)   ;
%Straight Lines [id:da6416258301598567] 
\draw    (619.52,210) -- (619.52,192) ;
\draw [shift={(619.52,190)}, rotate = 90] [color={rgb, 255:red, 0; green, 0; blue, 0 }  ][line width=0.75]    (10.93,-3.29) .. controls (6.95,-1.4) and (3.31,-0.3) .. (0,0) .. controls (3.31,0.3) and (6.95,1.4) .. (10.93,3.29)   ;
%Straight Lines [id:da6852878208400004] 
\draw  [dash pattern={on 0.84pt off 2.51pt}]  (507,172) -- (526.52,172) -- (539.52,172) ;
%Straight Lines [id:da02487290612432025] 
\draw    (419.52,210) -- (419.52,192) ;
\draw [shift={(419.52,190)}, rotate = 90] [color={rgb, 255:red, 0; green, 0; blue, 0 }  ][line width=0.75]    (10.93,-3.29) .. controls (6.95,-1.4) and (3.31,-0.3) .. (0,0) .. controls (3.31,0.3) and (6.95,1.4) .. (10.93,3.29)   ;

% Text Node
\draw (221,162) node [anchor=north west][inner sep=0.75pt]   [align=left] {LSTM};
% Text Node
\draw (600,160) node [anchor=north west][inner sep=0.75pt]   [align=left] {LSTM};
% Text Node
\draw (400,161) node [anchor=north west][inner sep=0.75pt]   [align=left] {LSTM};
% Text Node
\draw (145,156) node [anchor=north west][inner sep=0.75pt]  [font=\scriptsize] [align=left] {a\textsubscript{-1}};
% Text Node
\draw (325,156) node [anchor=north west][inner sep=0.75pt]  [font=\scriptsize] [align=left] {a\textsubscript{0}};
% Text Node
\draw (704,155) node [anchor=north west][inner sep=0.75pt]  [font=\scriptsize] [align=left] {a\textsubscript{N}};
% Text Node
\draw (146,176) node [anchor=north west][inner sep=0.75pt]  [font=\scriptsize] [align=left] {c\textsubscript{-1}};
% Text Node
\draw (326,176) node [anchor=north west][inner sep=0.75pt]  [font=\scriptsize] [align=left] {c\textsubscript{0}};
% Text Node
\draw (705,174) node [anchor=north west][inner sep=0.75pt]  [font=\scriptsize] [align=left] {c\textsubscript{N}};
% Text Node
\draw (236,214) node [anchor=north west][inner sep=0.75pt]  [font=\scriptsize] [align=left] {x\textsubscript{0}};
% Text Node
\draw (415,214) node [anchor=north west][inner sep=0.75pt]  [font=\scriptsize] [align=left] {x\textsubscript{1}};
% Text Node
\draw (615,213) node [anchor=north west][inner sep=0.75pt]  [font=\scriptsize] [align=left] {x\textsubscript{N}};

\end{tikzpicture}
\end{center}

In summary, we discretize a sample path $X_t$ into $N$ pieces, and plug it into a neural network that consists of $N+1$ LSTM blocks, we get an output vector $\boldsymbol a = (a_0, a_1, ..., a_N)$. The next thing to do is use the output vector to model the value function $V(X_t)$.   

\subsection{Value Function Approximation}
To approximate the value function $v(X_t)$, where $t_i \leq t < t_{i+1}$, we use the most well-known neural network --- feed-forward neural network, or fully connected neural network with three inputs, $t_i$, $x_{i}$, and $a_{i-1}$, and one output, denote as $\phi (t_i, x_i, a_{i-1}; \theta^f)$, where $\theta^f$ means the parameters in the feed-forward network. The exact structure of the feed-forward network (the number of hidden layers, the width of each layer, and the exact activation function) will be discussed in the empirical studies, and won't be important for the rest of the section.  

To summarize the above procedure, denote $u(X_t ; \theta^f, \theta^{lstm}) = \phi (t_i, x_i, a_{i-1}; \theta^f)$, where $\theta^{lstm}$ denotes parameters of LSTM neural networks. The partial derivatives of the value function can thus be represented as (here, we adapt the notation in \cite{saporito2020pdgm})
\begin{align}
   & u(X_{t_i}^h ; \theta^f, \theta^{lstm}) = \phi (t_i, x_i + h, a_{i-1}; \theta^f) \\
   & u(X_{t_i, \delta t} ; \theta^f, \theta^{lstm}) = \phi (t_i, x_i, a_{i-1}; \theta^f) \\
   & \Delta^{[\delta t]}_t u(X_{t_i} ; \theta^f, \theta^{lstm}) = \frac{u(X_{t_i, \delta t} ; \theta^f, \theta^{lstm}) - u(X_{t_i} ; \theta^f, \theta^{lstm})}{\delta t} \\ 
   & \Delta^{[h]}_x u(X_{t_i} ; \theta^f, \theta^{lstm}) = \frac{u(X_{t_i}^h ; \theta^f, \theta^{lstm}) - u(X_{t_i} ; \theta^f, \theta^{lstm})}{h} \\
   & \Delta^{[h]}_{xx} u(X_{t_i} ; \theta^f, \theta^{lstm}) = \frac{u(X_{t_i}^h ; \theta^f, \theta^{lstm}) - 2u(X_{t_i} ; \theta^f, \theta^{lstm})  +  u(X_{t_i}^{-h} ; \theta^f, \theta^{lstm})}{h^2}
\end{align}

In this way, with the "right" parameters $\theta^f$, and $\theta^{lstm}$, we can approximate $V(X_t)$ by $u(X_t; \theta^f, \theta^{lstm})$. In order to achieve the "right" coefficients, we need to train the above model with simulated sample paths.

Consider the equation $(28)$, to simplify the notations, let 
\begin{align}
& A = \frac{|| \boldsymbol \sigma_t^{-1} (\boldsymbol \mu_t - r \boldsymbol e_d) ||^2 }{2} \\
& B = \frac{\gamma}{2}\big [ d - d\text{\hspace{0.07cm}ln\hspace{0.07cm}} 2\pi e \gamma + \text{ln\hspace{0.07cm}} \text{det} (\boldsymbol \sigma_t^{T} \boldsymbol \sigma_t) \big ] \\
& C = \frac{\gamma d}{2}
\end{align}
Thus, equation $(28)$ becomes 
\begin{equation}
    \Delta_t V(X_t) - A \frac{[\Delta_x V(X_t)]^2}{\Delta_{xx} V(X_t)} + C \hspace{0.07cm} \text{ln} \hspace{0.07cm} \Delta_{xx} V(X_t) + B = 0
\end{equation}
If there are $M$ simulated paths, and we discretize each sample path into $N$ pieces. Denote $X_t^{(j)}$ be the $j^{th}$ simulated path. To save the space, let $\theta = [\theta^f, \theta^{lstm}]$. then the loss function to minimize is 
\begin{align}
    J_{N,M}(\theta) &= \frac{1}{M} \frac{1}{N} \sum_{j = 1}^M \sum_{i = 0}^M \big( \Delta_t^{[\delta t]} u(X_t^{(j)}; \theta) - A \frac{[\Delta_x^{[h]} u(X_t^{(j)}; \theta)]^2}{\Delta_{xx}^{[h]} u(X_t^{(j)}; \theta)} + C \hspace{0.07cm} \text{ln} \hspace{0.07cm}\Delta_{xx}^{[h]} u(X_t^{(j)}; \theta) + B \big )^2 
\end{align}
Here, $\alpha$ is a hyper-parameter to be determined depending on the situation, the last term is in order to let the average of final values of the simulated paths be as close to the expected return as possible.

\section{Empirical Studies}
There are two phases of empirical studies, the first is the clustering phase, and the second is the portfolio construction phase. 

In the clustering phase, the simulated annealing clustering method with $\kappa = 0.0001$, $w = 0.5$, $\alpha = 0.99$, $T_0 = 100$, and $T_f = 0.1$ is used to cluster stocks in the S\&P 500 index based on 2020-2022 historic data into 25 groups. The procedure is repeated 100 times to minimize the energy function. Figures 1-4 show the cumulative returns of stocks from four different groups. The results indicate that most stocks in a group perform similarly, while there may be some exceptional stocks that perform differently. This finding suggests that the clustering method could be used for pair trading to identify statistical arbitrage opportunities if the size of each cluster is limited to 2-5 stocks. However, this research direction is not explored in this paper and left for future studies.

In the second phase, 25 stocks are randomly selected from each group and a mean-variance portfolio is constructed based on their historic data. Before constructing the portfolio, the drift rate and covariance matrix for each day must be estimated. In this paper, the estimation is based on the past 75 days of stock returns, which may not be the most accurate method. More accurate estimation can be achieved with high-frequency data and sufficient computation resources.

The parameters used in the trading strategy were set as follows: $\gamma = 0.01$, $d = 25$, and $r = 0$ since the interest rate has been zero for most of the past two years. The initial wealth path was $[1.0, 1.01]$ and the trading strategy began with 1 dollar. At the start of each period, the current wealth path $X_t$ was used to generate hundreds of simulated paths with the same drift rate and volatility. These simulated paths were then used to train the neural network, which produced the policy parameters for generating holdings for the next period. A new wealth point was added to the current wealth path after each iteration, therefore, a new neural network needs to be trained. Due to limited computational resources, the trading strategy was only tested for 30 consecutive trading days (30 iterations) at three different periods covering bull, bear, and volatile markets. Figure 5-7 shows that this trading strategy outperformed the S\&P 500 index. 

\section{Conclusion}
This paper introduces a novel asset clustering method and extends the exploratory mean-variance framework to the path-dependent case. To further improve this framework, possible enhancements include replacing LSTM with transformer to enhance the neural network's structure, using high-frequency data to estimate model parameters, and adding constraints on leverage or portfolio asset percentages to enhance the framework's robustness.  

\newpage
\bibliography{Reference.bib}
\bibliographystyle{apacite}

\newpage

\begin{figure}[p]
    \centering
    \includegraphics[width = \textwidth]{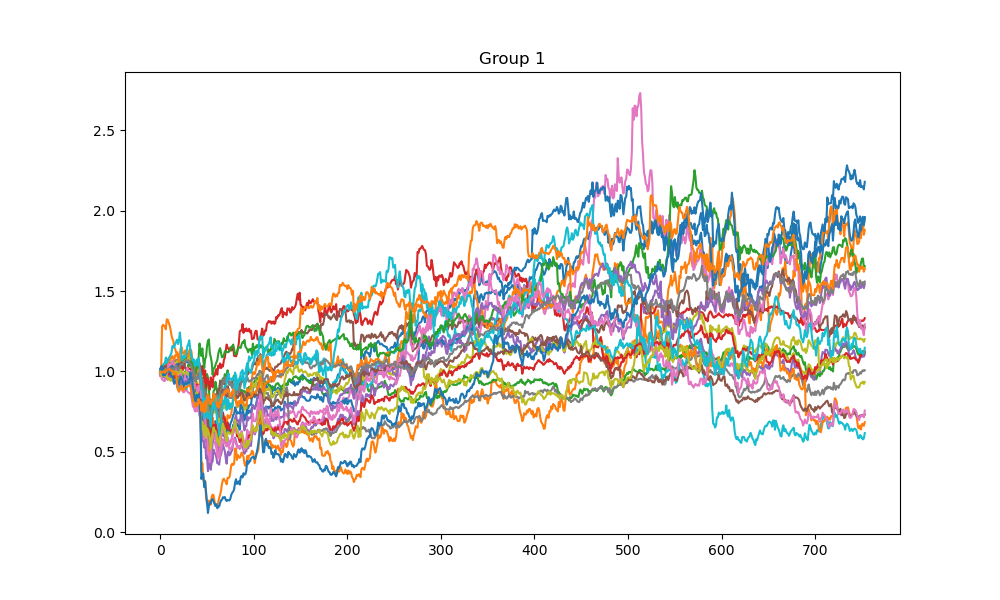}
    \caption{cumulative returns of group 1 stocks}
    \label{fig:my_label}
\end{figure}

\begin{figure}[p]
    \centering
    \includegraphics[width =  \textwidth]{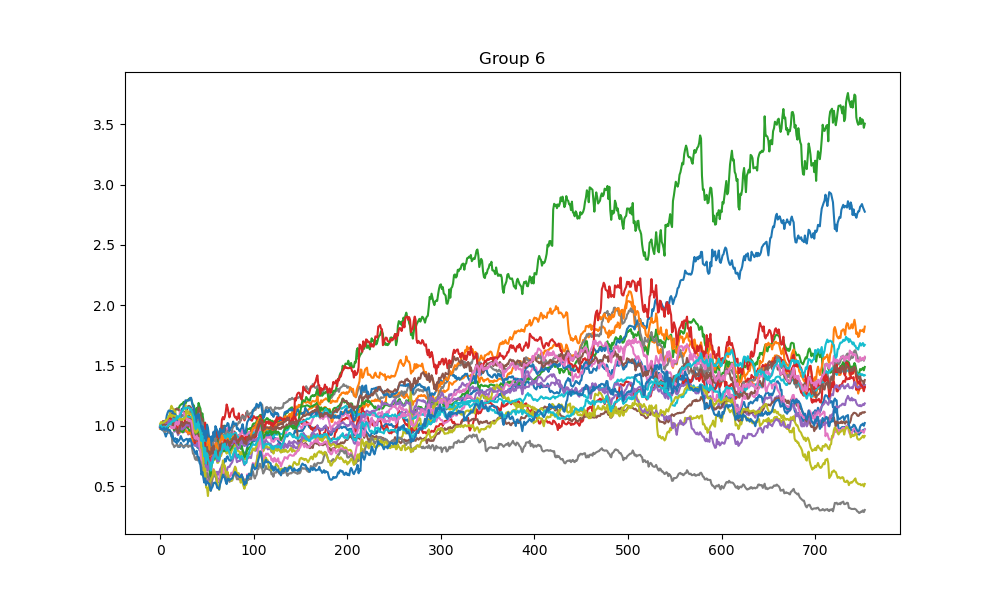}
    \caption{culmulative returns of group 6 stocks}
    \label{fig:my_label}
\end{figure}

\begin{figure}[p]
    \centering
    \includegraphics[width = \textwidth]{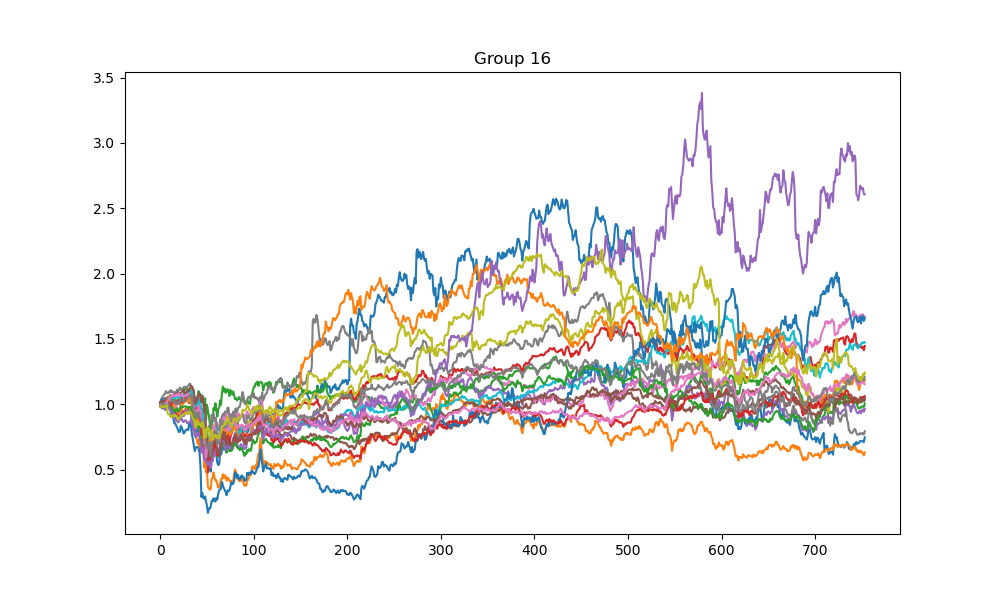}
    \caption{cumulative returns of group 16 stocks}
    \label{fig:my_label}
\end{figure}

\begin{figure}[p]
    \centering
    \includegraphics[width = \textwidth]{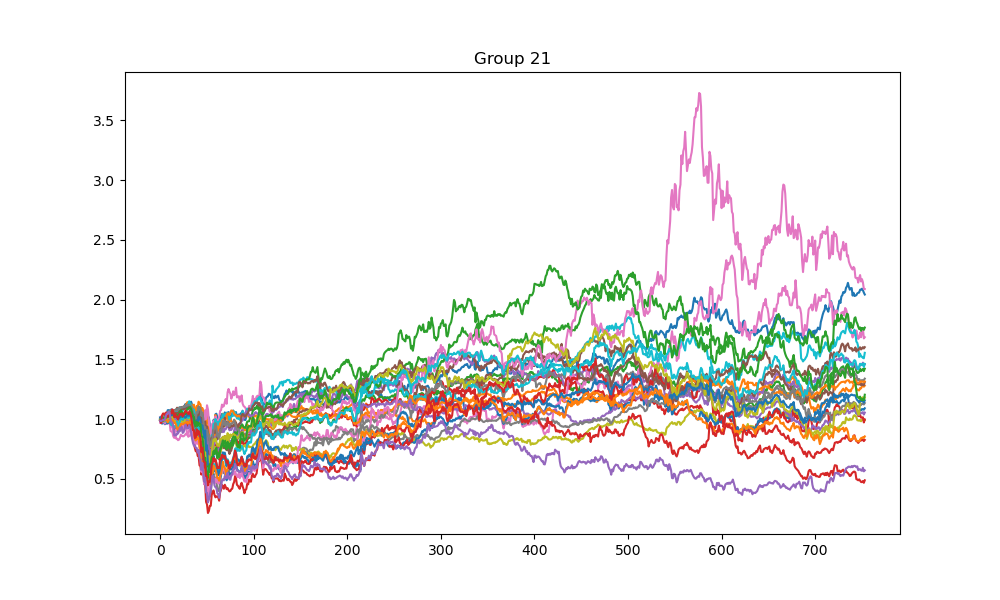}
    \caption{cumulative returns of group 21 stocks}
    \label{fig:my_label}
\end{figure}

\begin{figure}[p]
    \centering
    \includegraphics[width = 14cm]{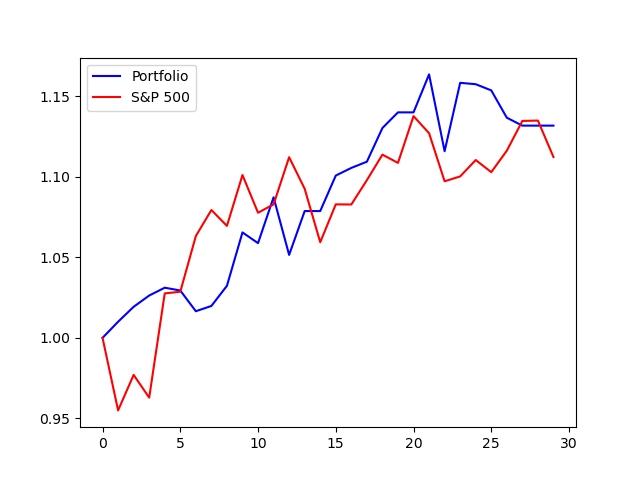}
    \caption{61 - 91 trading days since 2020-01-01}
    \label{fig:my_label}
\end{figure}

\begin{figure}[p]
    \centering
    \includegraphics[width = 14cm]{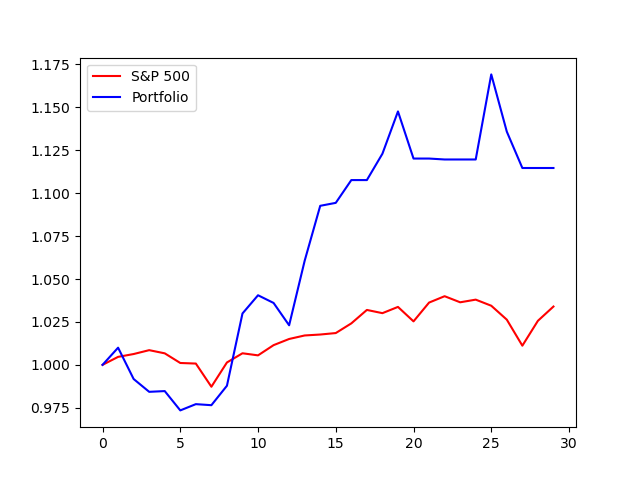}
    \caption{361 - 391 trading days since 2020-01-01}
    \label{fig:my_label}
\end{figure}

\begin{figure}[p]
    \centering
    \includegraphics[width = 14cm]{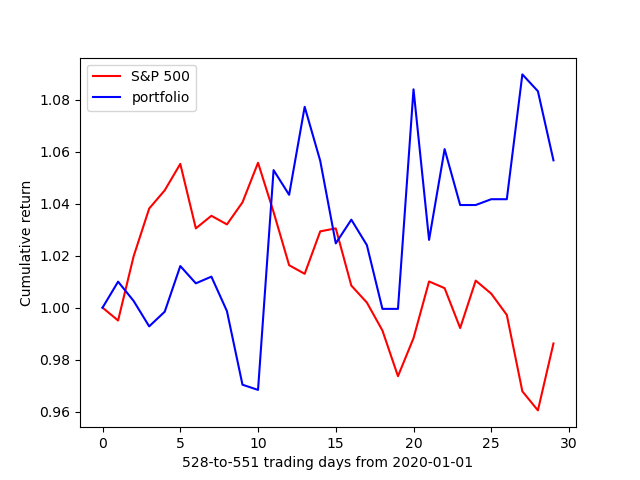}
    \caption{521 - 551 tradings days since 2020-01-01}
    \label{fig:my_label}
\end{figure}

\end{document}